\documentclass[final,5p,times,twocolumn]{elsarticle}
\usepackage{amsmath,amssymb}
\usepackage{array}
\usepackage{graphicx}
\usepackage{dcolumn}
\usepackage{bm}
\usepackage{multirow}
\usepackage{hyperref}
\hypersetup{
    colorlinks = true,
    linkcolor = blue,
    citecolor = blue,
    urlcolor = blue,
}
\journal{Physics Letters B}
\biboptions{sort&compress}

\begin{document}
\begin{frontmatter}
\title{Thermodynamic topological classification of magnetically charged slowly rotating Kerr black holes in nonlinear electrodynamics with a cosmological constant}

\author[gzu]{Peng Zhao}
\author[gzu]{Yu-Die Wan}
\author[gzu]{Zheng-Wen Long\corref{cor1}}
\ead{zwlong@gzu.edu.cn}
\cortext[cor1]{Corresponding author}
\affiliation[gzu]{organization={College of Physics, Guizhou University},
  city={Guiyang},
  postcode={550025},
  state={Guizhou},
  country={People's Republic of China}}

\begin{abstract}
We provide an explicit rotating black-hole realization of the predicted \(\overline{W}^{1-}\) thermodynamic topological subclass by studying a slowly rotating, magnetically charged Kerr black hole in nonlinear electrodynamics (NLED). Using a reduced fixed-parameter off-shell potential, we derive the inverse-temperature curve analytically rather than relying solely on numerical plots. In AdS spacetime, this curve begins at a finite nonzero value, increases monotonically, and diverges at large horizon radius; it therefore contains exactly one defect with \(w=W=-1\) and has no stable large-black-hole branch. Positivity of the analytic coefficient establishes this result throughout the admissible slow-rotation parameter domain. Under the same reduced prescription, the Kerr-AdS benchmark instead has two branches and belongs to the \(W^{0-}\) class, establishing a clear topological distinction between the NLED-Kerr and classical rotating cases. In dS spacetime, the formally extended off-shell curve remains monotonic and any isolated zero has \(w=-1\); because the parameter-dependent horizon-admissible range is not fixed in the present reduced analysis, no physical dS asymptotic subclass is assigned.
\end{abstract}

\end{frontmatter}


\section{Introduction}
\label{sec:level1}
Black holes are central predictions of general relativity and serve as an important framework for probing the interplay among gravity, quantum mechanics, and thermodynamics\cite{ref1,ref2,ref3,ref4}. Bekenstein first proposed that the entropy of a black hole is proportional to the area of its event horizon, while Hawking, based on quantum field theory, demonstrated that black holes emit thermal radiation, endowing them with a well-defined temperature and a complete thermodynamic description, thus establishing black hole thermodynamics\cite{ref3,ref4}. In the anti-de Sitter (AdS) spacetime, black holes exhibit rich phase structures and critical behaviors\cite{ref7,ref8,ref9,ref10,ref11}. Hawking and Page discovered a first-order phase transition between the Schwarzschild-AdS black hole and thermal radiation\cite{ref12}, which, within the AdS/CFT correspondence, is interpreted as the confinement-deconfinement phase transition in gauge theories. By treating the cosmological constant as thermodynamic pressure and extending to the extended phase space, charged AdS black holes display a phase transition between small and large black holes that closely resembles the van der Waals fluid behavior \cite{ref13}. 

In recent years, topological methods have been introduced into black hole thermodynamics, providing a novel approach to uncovering the universal properties of black hole systems. Wei, Liu, et al. defined black hole solutions as topological defects in the thermodynamic parameter space\cite{ref14,ref15}. By constructing a topological vector field via the generalized off-shell Helmholtz free energy and employing Duan's \(\phi\)-mapping theory, they defined the winding number and the global topological charge, thereby accomplishing a topological classification of black hole thermodynamics\cite{ref16,ref17,ref18,ref19,ref20}. In this framework, the sign of the winding number directly reflects the local thermodynamic stability: a positive winding number signifies a stable thermodynamic branch, while a negative one corresponds to an unstable branch; the sum of all local winding numbers yields the global topological charge of the system. Furthermore, by incorporating the asymptotic behavior of the inverse temperature at the extremal horizon radius and the stability characteristics of the thermodynamic branches, black holes can be categorized into several universal topological classes\cite{ref21}. To date, this classification scheme has been applied to a wide range of classical black hole models, and studies have shown that the cosmological constant, electric charge, angular momentum, and spacetime dimension can significantly modify the topological structure of black hole thermodynamics\cite{ref22,ref23,ref24,ref25,ref26}. Further examples include topological charged dilatonic black holes and black holes in \(F(R)\)-ModMax gravity's rainbow, for which the thermodynamic topology can depend on the statistical ensemble, horizon topology, nonlinear electromagnetic sector, and modified-gravity parameters \cite{refb1,refb2}. In five-dimensional charged Gauss--Bonnet AdS black holes, the Gauss--Bonnet coupling was also shown to modify the thermodynamic topological classification in the grand-canonical ensemble relative to the corresponding RN--AdS case \cite{refb3}. Ensemble dependence has likewise been reported for phantom AdS black holes in massive gravity and charged dRGT black strings \cite{refPhantomMassive2024,refDRGTString2026}. In other settings, a conformal Killing gravity parameter can change the universal topological class \cite{refCKG2025}, whereas a charged AdS black hole with an antisymmetric-tensor background was found to retain an RN--AdS-like thermodynamic defect topology in the setup studied in Ref.\cite{refTensorAdS2026}.There is a general consensus that the AdS boundary tends to stabilize the large-radius black hole branch, thereby altering the global topological type of the system. One of the central questions in this field is whether additional physical corrections can break this stabilization mechanism induced by the AdS background. Recent developments have significantly extended the topological classification framework, revealing that the four originally established classes are insufficient to accommodate all black hole solutions. Subsequently, one novel topological class $W^{0-\longleftrightarrow 1+}$
and two new subclasses $\overline{W}^{1+}$ and $\hat{W}^{1+}$ were identified in gauged supergravity and dyonic black holes \cite{ref27}, while the $\tilde{W}^{1+}$ subclass was discovered in higher odd-dimensional multiply rotating Kerr-AdS black holes \cite{ref28}. Very recently, based on symmetry arguments of the inverse temperature functions, a new subclass $\overline{W}^{1-}$ was predicted to exist \cite{ref27}, though its physical realization in explicit black hole solutions had not yet been established. This raises a compelling question: can concrete black hole models realize this predicted topological subclass? 

Nonlinear electrodynamics (NLED), as a generalization of Maxwell's classical electrodynamics, provides an ideal platform for investigating the aforementioned issues \cite{ref29,ref30,ref31,ref32,ref33}. Born and Infeld originally proposed the NLED model to eliminate the divergence of the self-energy of point charges \cite{ref34}; when coupled to general relativity, it allows for the construction of regular black hole solutions free from central singularities \cite{ref29,ref33,ref36}, and has found wide applications in cosmology, dark energy, and other fields \cite{ref38,ref39}. In contrast to the linear electromagnetic field, NLED modifies the near-horizon geometry and small-scale structure of black holes. Recent studies of Born--Infeld AdS black holes surrounded by perfect-fluid dark matter further show that nonlinear electromagnetic and environmental parameters can jointly modify the horizon structure, thermodynamic stability, and critical behavior \cite{refa}. For magnetically charged NLED black holes, the conventional constant mass is replaced by an effective mass function that varies smoothly with the radial coordinate. This modification directly alters the black hole temperature, heat capacity, and the morphology of thermodynamic branches, thereby modifying the asymptotic behavior of the inverse temperature, which is a key ingredient in thermodynamic topology. In particular, the effective mass function induced by NLED may provide the missing ingredient necessary to realize the recently predicted $\overline{W}^{1-}$ topological subclass, characterized by one unstable large-black-hole branch at low temperature, no black-hole state in the high-temperature limit, and a global topological number $W=-1$. 

In this work, we adopt the NLED model proposed by Kruglov, which reduces naturally to Maxwell's theory in the weak-field limit and admits regular magnetically charged black hole solutions with finite total mass \cite{ref40}. By incorporating the slow-rotation approximation, we construct a slowly rotating Kerr black hole with magnetic charge in dS/AdS spacetime within this NLED framework. Although the fundamental thermodynamic quantities of this black hole, such as the metric, temperature, and entropy, have been derived previously, its thermodynamic topological properties remain largely unexplored. In particular, whether the nonlinear electromagnetic correction can suppress the stabilizing effect of the AdS background on the large-branch black hole remains an open question. 

Motivated by these open issues, we study the NLED-Kerr model with a reduced fixed-parameter off-shell potential. This restriction is stated explicitly and is not identified with a strict grand-canonical ensemble. In AdS spacetime, the analytic defect curve has a finite nonzero lower endpoint, diverges at infinity, and possesses one zero with $w=-1$, thereby realizing the predicted $\overline{W}^{1-}$ subclass. In dS spacetime, the formally extended off-shell curve is monotonic and each isolated zero has $w=-1$; because the parameter-dependent horizon-admissible range is not fixed in the present reduced analysis, no physical dS asymptotic subclass is assigned. The monotonicity proof applies throughout the admissible domain of \(a\), \(\beta\), and \(q_m\), while the figures illustrate representative parameter choices. Comparison with Kerr-AdS is made under the same fixed-rotation reduction and shows a different endpoint and branch structure.

The remainder of this paper is organized as follows. Section II reviews the thermodynamic-topology formalism and the relevant topological classes. Section III introduces the magnetically charged slowly rotating NLED-Kerr model, analyzes its dS and AdS sectors, and compares the result with Kerr-AdS. Section IV summarizes the conclusions and their domain of validity.
\section{Basics of thermodynamic topology }
\label{sec:level2}
In this section, we briefly review the thermodynamic-topology formalism and the established and predicted topological (sub)classes needed for the subsequent analysis.

We regard the thermodynamic states of black holes as intrinsic topological defects embedded in the extended thermodynamic parameter manifold. To implement a topological classification for different black holes, we introduce a generalized off-shell Helmholtz free energy functional \cite{ref41,ref42,ref43,ref44}
\begin{equation}
    \mathcal{F} = M - \frac{S}{\tau},
    \label{eq:ziyouneng}
\end{equation}
where \(\tau\) denotes the external inverse-temperature parameter of the cavity, which allows the system to be taken off shell. When \(\tau\) equals the inverse Hawking temperature, i.e., \(\tau=1/T\), the generalized free energy reduces to the standard on-shell Helmholtz free energy, \(F=M-TS\) \cite{ref45}.

To systematically investigate the topological defect structure of black hole thermodynamics, we construct a two-dimensional vector field in the parameter space spanned by the event horizon radius \(r_h\) and an auxiliary angular coordinate \(\Theta\) \cite{ref41}
\begin{equation}
    \phi = \left( \phi^{r_h}, \phi^\Theta \right) = \left( \frac{\partial \hat{\mathcal{F}}}{\partial r_h}, \frac{\partial \hat{\mathcal{F}}}{\partial \Theta} \right).
    \label{eq:fai}
\end{equation}
Here the angularly modified free energy is defined by \(\hat{\mathcal{F}}=\mathcal{F}+\csc\Theta\), so that \(\phi^\Theta=-\cot\Theta\csc\Theta\). The auxiliary coordinate satisfies \(\Theta\in(0,\pi)\); the vector field is singular only at the two angular boundaries. The isolated zeros of \(\phi\) correspond to on-shell black-hole equilibrium states and constitute the topological defects of the thermodynamic parameter space. Their winding numbers distinguish locally stable and unstable branches.

Motivated by Ref. \cite{ref46}, the topological current is defined as
\begin{equation}
J^{\mu}=\frac{1}{2 \pi} \epsilon^{\mu \nu \lambda} \epsilon_{A B} \partial_{\nu} n^{A} \partial_{\lambda} n^{B},
\end{equation}
where \(x^\mu=(\tau,r_h,\Theta)\), \(n^A=\phi^A/\|\phi\|\), and \(A,B=1,2\) label the two internal components of the vector field. Capital internal indices are used here to avoid confusion with the rotation parameter \(a\). This current obeys the conservation law \(\partial_\mu J^\mu=0\). The topological charge over a parameter region \(\Sigma\) is then defined as
\begin{equation}
W\equiv Q_{t}=\int_{\Sigma} J^{0} d^{2} x=\sum_{i=1}^{N} w_{i},
\end{equation}
where \(w_i\) is the winding number of the \(i\)th isolated zero of \(\phi^A\). The global topological number \(W\) is the sum of all local winding numbers in \(\Sigma\).

This theoretical framework provides an effective tool for classifying different black holes. The classes and subclasses relevant to the present discussion are listed below; for further details, see Refs. \cite{ref14,ref27,ref28,ref47}.
\begin{equation}
\begin{gathered}
W^{1-},\ W^{0+},\ W^{0-},\ W^{1+},\ \bar{W}^{1+},\ \hat{W}^{1+},\\
\widetilde{W}^{1+},\ W^{0-\leftrightarrow 1+},\ \ddot{W}^{1-},\ \overline{W}^{1-}.
\end{gathered}
\end{equation}
Because \(\tau\) is the inverse-temperature parameter, the endpoint behaviors relevant to the present comparison can be summarised as
\begin{align*}
W^{0+}&: \quad \tau(r_{m}) = \infty,\ \tau(\infty) = \infty, \\
W^{1-}&: \quad \tau(r_{m}) = 0,\ \tau(\infty) = \infty, \\
W^{0-}&: \quad \tau(r_{m}) = 0,\ \tau(\infty) = 0, \\
W^{1+}&: \quad \tau(r_{m}) = \infty,\ \tau(\infty) = 0, \\
\overline{W}^{1-}&: \quad \tau(r_{m}) = \text{finite and nonzero},\ \tau(\infty) = \infty.
\end{align*}
\begin{table*}[!htbp]
\centering
\caption{Thermodynamic properties of representative established and predicted topological (sub)classes, including
$W^{1-}$, $W^{0+}$, $W^{0-}$, $W^{1+}$\cite{ref14}, $W^{0-\leftrightarrow 1+}$, $\bar{W}^{1+}$, $\hat{W}^{1+}$\cite{ref27}, $\widetilde{W}^{1+}$\cite{ref28}, $\ddot{W}^{1-}$\cite{ref47}, and the $\overline{W}^{1-}$ subclass studied here.} 
\label{tab:1}   
\large
\renewcommand{\arraystretch}{1.3}
\resizebox{1\textwidth}{!}{%
\begin{tabular}{|c|c|c|c|c|c|c|}\hline

Topological (sub)classes & Innermost & Outermost & Low $T$ ($\tau\to\infty$) & High $T$ ($\tau\to0$) & DP & $W$ \\\hline

$W^{1-}$               & Unstable & Unstable & Unstable large & Unstable small & In pairs & $-1$ \\\hline
$W^{0+}$               & Stable   & Unstable & Unstable large + stable small & No & One more GP & $0$ \\\hline
$W^{0-}$               & Unstable & Stable   & No & Unstable small + stable large & One more AP & $0$ \\\hline
$W^{1+}$               & Stable   & Stable   & Stable small & Stable large & In pairs & $+1$ \\\hline
$W^{0-\leftrightarrow 1+}$ & Unstable & Stable & No & Stable large & One more AP & $0$ or $+1$ \\\hline
$\bar{W}^{1+}$         & Stable   & Stable   & No & Stable large & In pairs & $+1$ \\\hline
$\hat{W}^{1+}$         & Stable   & Stable   & Unstable small + two stable small & Stable large & One more GP & $+1$ \\\hline
$\widetilde{W}^{1+}$   & Unstable & Stable   & Stable small & Unstable small + stable small + stable large & One more AP & $+1$ \\\hline
$\ddot{W}^{1-}$& Unstable & Stable   & Unstable small & Unstable small + unstable small + stable large & One more AP & $-1$ \\ \hline
$\overline{W}^{1-}$& Unstable & Unstable & Unstable large & No & In pairs & $-1$ \\ \hline

\end{tabular}%
}
\end{table*}

Table \ref{tab:1} summarizes the endpoint stability, low- and high-temperature states, defect processes, and global topological numbers of the representative classes. Here DP denotes the defect process, while GP and AP denote generation and annihilation points, respectively.

Following the generalized off-shell free-energy construction and its applications to rotating black holes \cite{ref41,refd1,refd2,refd3,refd4}, we introduce the following one-dimensional restriction along a path with \(\lambda=(a,q_m,\beta,L)\) held fixed:
\begin{equation}
\begin{aligned}
\mathcal{F}_{\mathrm{red}}(r_h;\tau,\lambda)
&=\mathcal{E}_{\mathrm{red}}(r_h;\lambda)
-\frac{S(r_h;\lambda)}{\tau},\\
\lambda&=(a,q_m,\beta,L).
\end{aligned}
\end{equation}
Here \(\mathcal{E}_{\mathrm{red}}\) is the explicit effective energy along the fixed-\(\lambda\) path given below. Thus, \(\mathcal{F}_{\mathrm{red}}\) is defined directly as a reduced path functional, not as a Legendre transform with independently varied angular or magnetic potentials. All NLED-Kerr and Kerr-AdS results below use the same fixed-rotation reduction.
\section{Slowly rotating Kerr--(A)dS black hole model with magnetic charge in nonlinear electrodynamics }
\label{sec:level3}
In this section, we introduce a slowly rotating Kerr black hole coupled to a cosmological constant in Kruglov's nonlinear electrodynamics (NLED), together with the metric, effective mass distribution, and thermodynamic quantities required for the topological analysis. Throughout this paper, we adopt natural units \(c=\hbar=G=1\) and work in the admissible parameter domain
\begin{equation}
\begin{aligned}
\mathcal{D}=\{&L>0,\ q_m>0,\ \beta>0,\ \Xi>0;\\
&0<|a|/L\leq0.1\}.
\end{aligned}
\label{eq:domain}
\end{equation}
The static boundary \(a=0\), the uncharged boundary \(q_m=0\), and the Maxwell boundary \(\beta=0\) are different reduced sectors and are not obtained by taking limits inside \(\mathcal{D}\). The reference length \(r_0\) is used only to normalize the vector-field plots.

We use the resummed slow-rotation metric ansatz reported for the source model and restrict every conclusion to Eq.~\eqref{eq:domain}. The retained higher powers of \(a\) are part of that effective ansatz; no extrapolation beyond \(|a|/L=0.1\), and no continuity claim at \(a=0\), is made.
\begin{equation}
\begin{aligned}
ds^2={}&-\frac{\Delta_r}{\rho^2}\left(dt-\frac{a\sin^2\theta}{\Xi}d\phi\right)^2
+\frac{\rho^2}{\Delta_r}dr^2\\
&+\frac{\rho^2}{\Delta_\theta}d\theta^2
+\frac{\Delta_\theta\sin^2\theta}{\rho^2}
\left(a\,dt-\frac{r^2+a^2}{\Xi}d\phi\right)^2,
\end{aligned}
\end{equation}
with
\begin{equation}
\begin{aligned}
\rho^2 &= r^2+a^2\cos^2\theta,\qquad
\Delta_\theta=1+\frac{a^2\Lambda}{3}\cos^2\theta,\\
\Xi&=1+\frac{a^2\Lambda}{3}.
\end{aligned}
\end{equation}

A distinctive feature of this model is that the point-mass source is replaced by a smooth effective mass distribution induced by the magnetic charge \(q_m\) and the nonlinearity parameter \(\beta\). Following the work of Pawar and Das \cite{ref48}, the effective mass distribution arising from the NLED contribution is
\begin{equation}
M(r)=\frac{q_m^{3/2}}{2\beta^{1/4}} \arctan\left(\frac{r}{\sqrt{q_m}\beta^{1/4}}\right).
\end{equation}
This mass distribution possesses a well-defined limiting behavior. As the radial coordinate tends to infinity, the arctangent function approaches its asymptotic value $\pi/2$, yielding a finite total mass: 
\begin{equation}
\lim_{r\to\infty} M(r)=\frac{\pi}{4}\frac{q_m^{3/2}}{\beta^{1/4}}.
\end{equation}
Thus, the effective mass grows continuously with the radial coordinate \(r\) and approaches a finite value at infinity. This behavior reflects the regularizing effect of NLED on the matter source and distinguishes the model from the conventional Kerr solution. In the subsequent analysis, we adopt this expression as the total mass function, corresponding to the \(m_0=0\) sector of the more general mass profile. A claim of complete spacetime regularity would additionally require finite curvature invariants and is not inferred from the mass limit alone.

The horizon structure is governed by the function 
\begin{equation}
\Delta_r = (r^2 + a^2)\left(1 - \frac{\Lambda}{3}r^2\right) - 2M(r)r,
\end{equation}
whose positive zeros determine the horizon radii. Depending on the parameters, a dS background \((\Lambda>0)\) may contain Cauchy, event, and cosmological horizons, whereas an AdS background \((\Lambda<0)\) may contain inner and outer black-hole horizons. Multi-horizon regular black holes have also been investigated through generalized holographic-complexity observables, which can distinguish different interior regions of the black-hole spacetime \cite{refc}. In each case, \(r_h\) denotes the event-horizon radius used in the thermodynamic analysis.

The thermodynamic quantities evaluated on these horizons are given by the Bekenstein--Hawking entropy, the Hawking temperature, and the angular velocity, respectively:
\begin{equation}
\begin{aligned}
S&=\frac{\pi(r_h^2+a^2)}{1+a^2\Lambda/3},\qquad
T=\frac{\Delta_r'(r_h)}{4\pi(r_h^2+a^2)},\\
\Omega_h&=\frac{a(1+a^2\Lambda/3)}{r_h^2+a^2}.
\end{aligned}
\end{equation}
These geometric and thermodynamic quantities provide the input required for constructing the off-shell free energy and its topological vector field.
\subsection{\label{sec:leve3A} Reduced fixed-parameter topology of NLED-Kerr black holes in dS spacetime}
We first apply the reduced prescription to the slowly rotating, magnetically charged NLED-Kerr model in a de Sitter (dS) background.

For dS spacetime, \(\Lambda=3/L^2\) and \(\Xi=1+a^2/L^2\). We consider the reduced off-shell coordinate \(r_h>0\), with lower endpoint \(r_m=0\). Since the horizon-admissible range depends on the positive roots of \(\Delta_r\) for each parameter set and is not fixed in the present reduced analysis, the curves below are interpreted as formal off-shell continuations. Using the reduced prescription introduced above, we write
\begin{align}
\mathcal{F}_{\mathrm{red}}
&= -\frac{\pi q_m^{3/2}\left(2 a^4 + 3 a^2 L^2 + L^2 r_h^2\right)}
{8 L^2 \left(a^2 + r_h^2\right) \beta^{1/4}} \nonumber\\
&\quad-\frac{L^2 \pi \left(a^2 + r_h^2\right)}
{\left(a^2 + L^2\right)\tau}.
\label{eq:fred_ds}
\end{align}
Here \(\tau\) is the external inverse-temperature parameter. Equation~\eqref{eq:fred_ds} is the defining reduced path functional: \(a,q_m,\beta,L\) are held fixed before taking \(\partial\mathcal{F}_{\mathrm{red}}/\partial r_h\).

According to Eq.~\eqref{eq:fai}, we adopt a two-component vector field
\begin{equation}
\phi^{r_h} = \frac{a^2\left(a^2+L^2\right)\pi q_m^{3/2} r_h}{2 L^2\left(a^2+r_h^2\right)^2 \beta^{1/4}} -\frac{2 L^2 \pi r_h}{\left(a^2+L^2\right)\tau},
\end{equation}
\begin{equation}
\phi^{\Theta} = -\cot \Theta \csc \Theta.
\end{equation}
Taking the boundary condition \(\phi^{r_h}=0\), the inverse temperature parameter \(\tau\) can be expressed as 
\begin{equation}
   \tau = \frac{4 L^4 \left(a^2+r_h^2\right)^2 \beta^{1/4}}{a^2\left(a^2+L^2\right)^2 q_m^{3/2}}.
\label{eq:tau_ds}
\end{equation}
For \(r_h>0\), its derivative is
\begin{equation}
\frac{d\tau}{dr_h}=\frac{16L^4\beta^{1/4}r_h(a^2+r_h^2)}{a^2(a^2+L^2)^2q_m^{3/2}}>0,
\end{equation}
which proves that the defect curve is monotonic. Its endpoint behavior is
\begin{equation}
\lim_{r_h \to 0} \tau(r_h) = \frac{4 a^2 L^4 \beta^{1/4}}{\left(a^2+L^2\right)^2 q_m^{3/2}},\quad \lim_{r_h \to \infty} \tau(r_h) = \infty.
\label{eq:14}
\end{equation}
Thus, the extended curve has the formal endpoint pattern of \(\overline{W}^{1-}\). Because \(d\tau/dr_h>0\), a fixed \(\tau\) gives exactly one zero on the extended off-shell domain when
\begin{equation}
\tau>\tau(0).
\label{eq:dsphysical}
\end{equation}
Whether this zero lies in a physical horizon interval must be determined separately from the roots of \(\Delta_r\) for each parameter set. Accordingly, only the formal local winding content is used below, while the physical asymptotic \(\overline{W}^{1-}\) label is reserved for the unbounded AdS domain.
\begin{figure}[!htbp]
  \centering
  \includegraphics[width=0.3\textwidth]{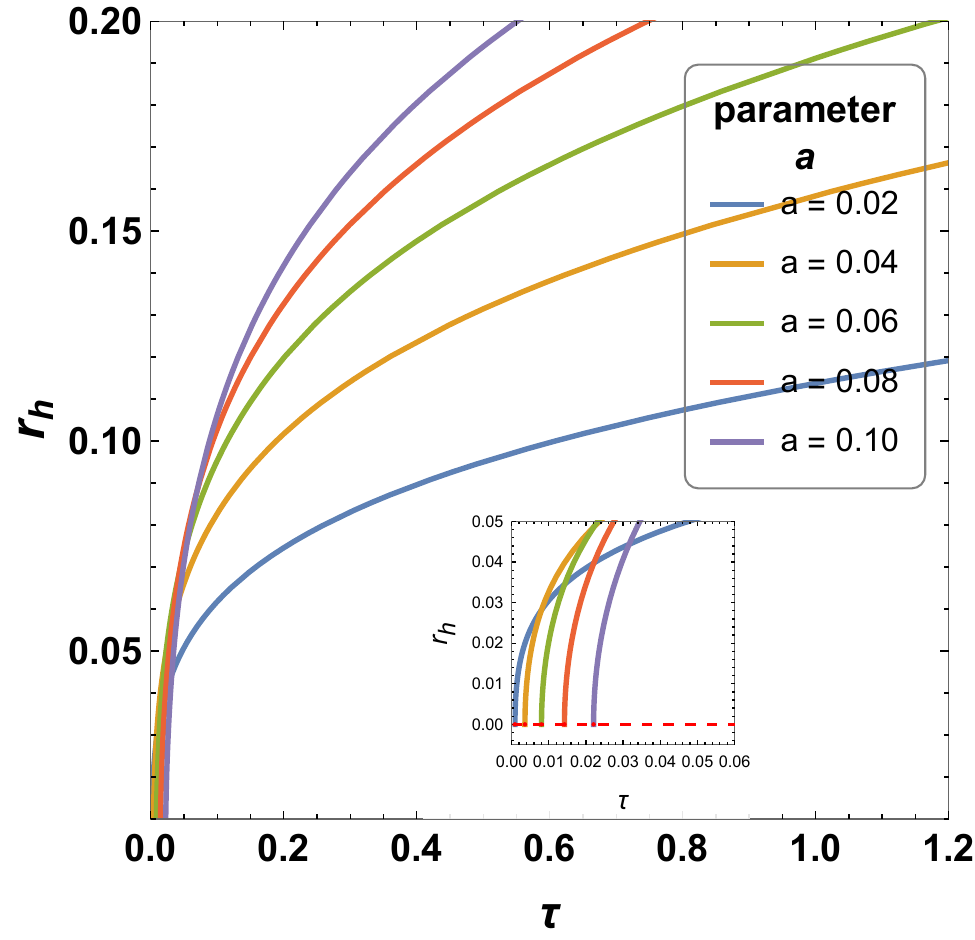}
  \caption{Formal reduced off-shell inverse-temperature curves \(\tau(r_h)\) for the NLED-Kerr model in dS spacetime, with \(L=1\), \(q_m=1\), \(\beta=1\), and \(a=0.02,0.04,0.06,0.08,0.10\).}
  \label{fig:jzzds_abianhua} 
\end{figure}
\begin{figure}[!htbp]
  \centering
  \includegraphics[width=0.3\textwidth]{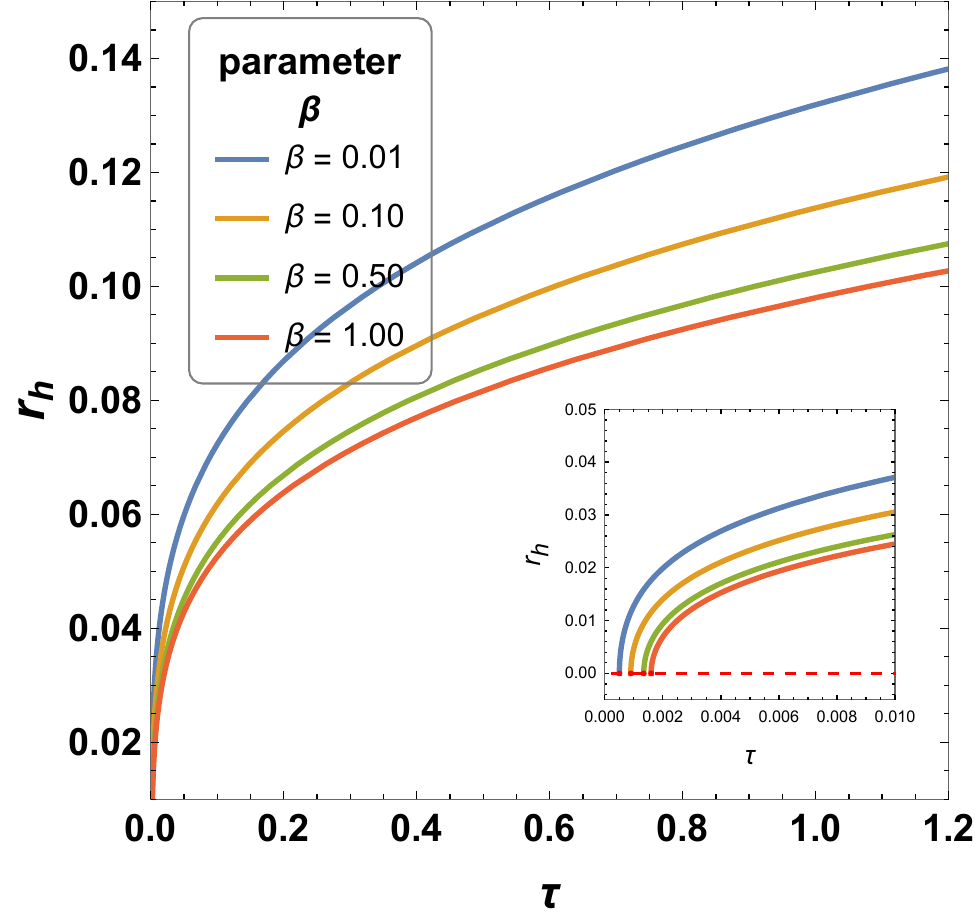}
  \caption{Formal reduced off-shell inverse-temperature curves \(\tau(r_h)\) for the NLED-Kerr model in dS spacetime, with \(L=1\), \(q_m=1\), \(a=0.02\), and \(\beta=0.01,0.1,0.5,1.0\).}
  \label{fig:jzzds_bbianhua} 
\end{figure}
\begin{figure}[!htbp]
  \centering
  \includegraphics[width=0.3\textwidth]{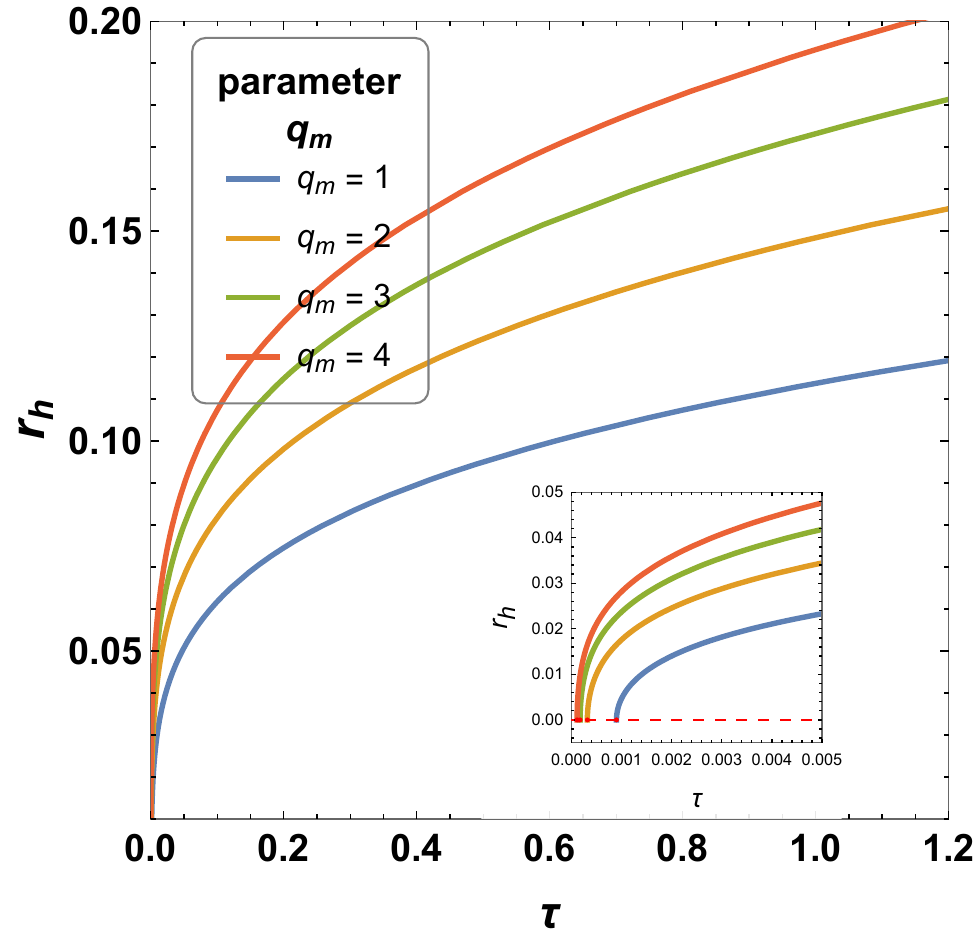}
  \caption{Formal reduced off-shell inverse-temperature curves \(\tau(r_h)\) for the NLED-Kerr model in dS spacetime, with \(L=1\), \(\beta=0.1\), \(a=0.02\), and \(q_m=1,2,3,4\).}
  \label{fig:jzzds_qbianhua} 
\end{figure}

Fig.~\ref{fig:jzzds_abianhua} illustrates the dS curves for several values of \(a\). Their common monotonicity is analytic, not numerical: the coefficient in Eq.~\eqref{eq:tau_ds} is positive everywhere in \(\mathcal{D}\), so \(d\tau/dr_h>0\) for every \(r_h>0\). The inset shows the finite positive lower-endpoint value.

Fig.~\ref{fig:jzzds_bbianhua} varies \(\beta\) at fixed \(L,q_m,a\). Since \(\tau\propto\beta^{1/4}\), increasing \(\beta\) rescales the curve without changing monotonicity or the sign of a zero.

Fig.~\ref{fig:jzzds_qbianhua} varies \(q_m\) at fixed \(L,\beta,a\). Since \(\tau\propto q_m^{-3/2}\), increasing \(q_m\) rescales the curve downward, again without changing monotonicity or the winding sign.
\begin{figure}[!htbp]
  \centering
  \includegraphics[width=0.3\textwidth]{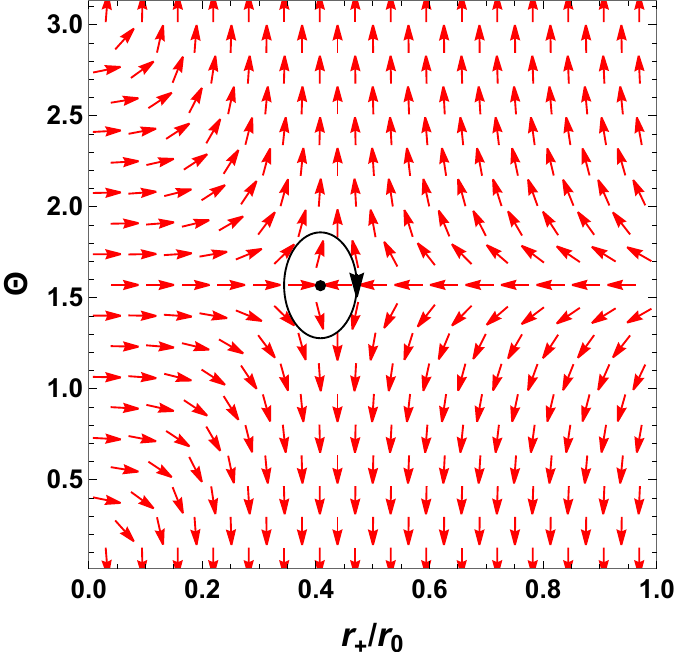}
  \caption{Formal reduced off-shell normalized vector field \(n^A\) in the \((r_h/r_0,\Theta)\) plane for the NLED-Kerr model in dS spacetime, with \(L=1\), \(q_m=1\), \(\beta=1\), \(a=0.05\), and \(\tau/r_0=60\). The black dot denotes the isolated zero.}
  \label{fig:jzzds_shiliangtu} 
\end{figure}

Fig.~\ref{fig:jzzds_shiliangtu} illustrates the extended vector field for \(L=q_m=\beta=1\), \(a=0.05\), and \(\tau/r_0=60\). Equation~\eqref{eq:tau_ds} gives \(r_h/r_0=0.4378\), consistent with the black dot. At \(\Theta=\pi/2\), \(\partial_\Theta\phi^\Theta>0\). Moreover, implicit differentiation of \(\phi^{r_h}=0\) gives
\begin{equation}
\frac{d\tau}{dr_h}
=-\frac{\partial_{r_h}\phi^{r_h}}{\partial_\tau\phi^{r_h}},
\qquad \partial_\tau\phi^{r_h}>0,
\end{equation}
so \(d\tau/dr_h>0\) implies \(\partial_{r_h}\phi^{r_h}<0\). The Jacobian is therefore negative and every nonzero defect has \(w=-1\). On the extended off-shell domain, Eq.~\eqref{eq:dsphysical} consequently gives one formal defect with \(W=-1\). Whether this defect belongs to the physical horizon domain must be checked separately, and the extended dS curve is not used to assign a physical asymptotic subclass \cite{ref27}.
\subsection{\label{sec:leve3B}Reduced fixed-parameter topology of NLED-Kerr black holes in AdS spacetime}
We now turn to the AdS background, where \(\Lambda=-3/L^2\), \(\Xi=1-a^2/L^2>0\), and the off-shell radial domain is unbounded. The defect curve becomes
\begin{equation}
\tau = \dfrac{4 L^4 \left(a^2 + r_h^2\right)^2 \beta^{1/4}}{\left(a^3 - a L^2\right)^2 q_m^{3/2}}.
\label{eq:tau_ads}
\end{equation}
For \(a\neq0\) and \(\Xi>0\), i.e., \(L^2>a^2\), the derivative
\begin{equation}
\frac{d\tau}{dr_h}=\frac{16L^4\beta^{1/4}r_h(a^2+r_h^2)}{(a^3-aL^2)^2q_m^{3/2}}>0,
\end{equation}
shows that the AdS defect curve is also monotonic for \(r_h>0\). Its limiting behaviors are
\begin{equation}
\lim_{r_h\to 0}\tau(r_h)=\frac{4 a^4 L^4 \beta^{1/4}}{\left(a^3-a L^2\right)^2 q_m^{3/2}}, 
\end{equation}
\begin{equation}
\lim_{r_h\to\infty}\tau(r_h)\sim \frac{4 L^4 \beta^{1/4}\,r_h^4}{\left(a^3-a L^2\right)^2 q_m^{3/2}}\to\infty.
\end{equation}
\begin{figure}[!htbp]
  \centering
  \includegraphics[width=0.3\textwidth]{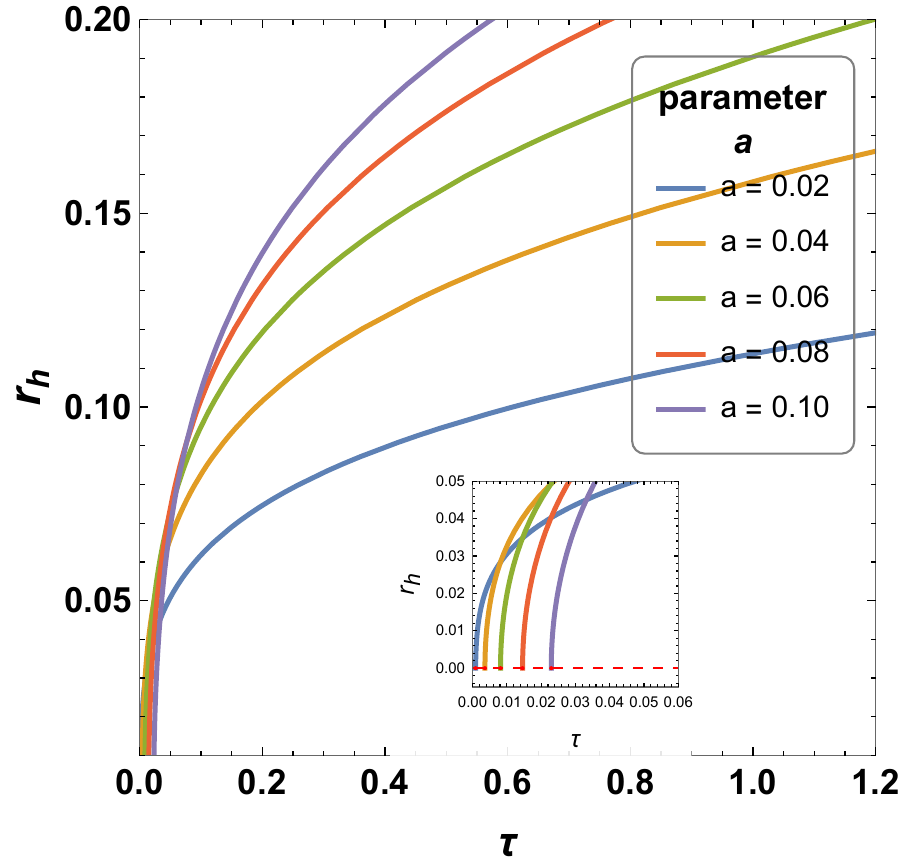}
  \caption{Inverse-temperature curves \(\tau(r_h)\) for the NLED-Kerr black hole in AdS spacetime, with \(L=1\), \(q_m=1\), \(\beta=1\), and \(a=0.02,0.04,0.06,0.08,0.10\).}
  \label{fig:jzzads_abianhua} 
\end{figure}
\begin{figure}[!htbp]
  \centering
  \includegraphics[width=0.3\textwidth]{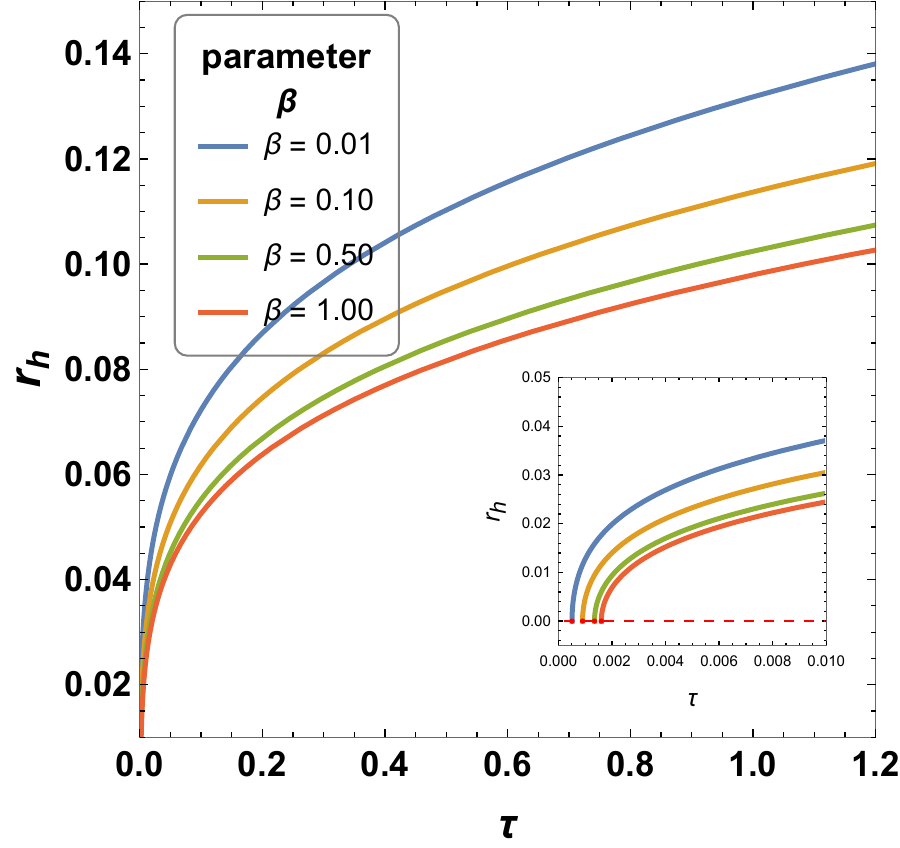}
  \caption{Inverse-temperature curves \(\tau(r_h)\) for the NLED-Kerr black hole in AdS spacetime, with \(L=1\), \(q_m=1\), \(a=0.02\), and \(\beta=0.01,0.1,0.5,1.0\).}
  \label{fig:jzzads_bbianhua} 
\end{figure}
\begin{figure}[!htbp]
  \centering
  \includegraphics[width=0.3\textwidth]{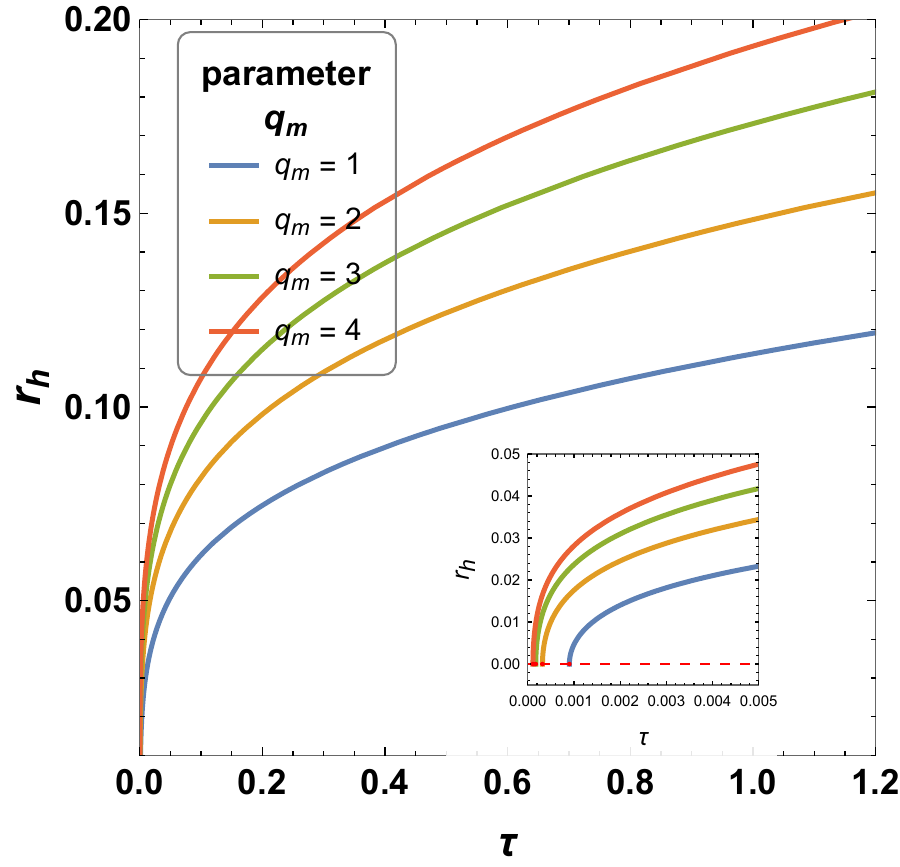}
  \caption{Inverse-temperature curves \(\tau(r_h)\) for the NLED-Kerr black hole in AdS spacetime, with \(L=1\), \(\beta=0.1\), \(a=0.02\), and \(q_m=1,2,3,4\).}
  \label{fig:jzzads_qbianhua} 
  \end{figure}  
The finite nonzero lower-endpoint value and the quartic divergence at infinity agree with the endpoint behavior of the \(\overline{W}^{1-}\) subclass. Because \((a^3-aL^2)^2>0\) throughout \(\mathcal{D}\), this endpoint type and strict monotonicity hold for every admissible \(a,q_m,\beta,L\), not only for the plotted values.

Fig.~\ref{fig:jzzads_abianhua} illustrates the analytic AdS result for several values of \(a\). Rotation changes the scale of \(\tau(r_h)\), but positivity of the coefficient and \(d\tau/dr_h>0\) preserve the finite lower endpoint, the divergence at infinity, and the \(\overline{W}^{1-}\) subclass throughout \(\mathcal{D}\).

Fig.~\ref{fig:jzzads_bbianhua} illustrates the scaling \(\tau\propto\beta^{1/4}\). Varying \(\beta>0\) rescales the curve but cannot change its monotonicity, endpoint types, or classification.

Fig.~\ref{fig:jzzads_qbianhua} illustrates the scaling \(\tau\propto q_m^{-3/2}\). Varying \(q_m>0\) likewise changes only the scale and leaves the \(\overline{W}^{1-}\) classification unchanged.

\begin{figure}[!htbp]
  \centering
  \includegraphics[width=0.3\textwidth]{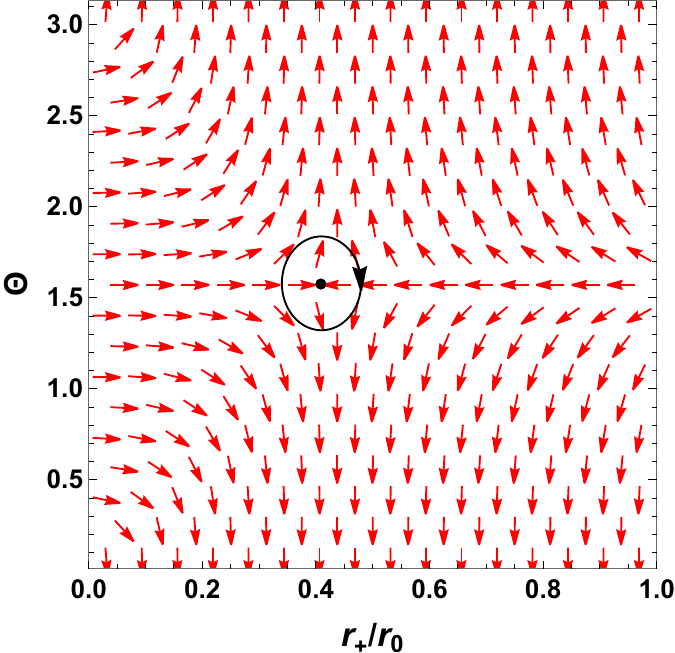}
  \caption{Normalized vector field \(n^A\) in the \((r_h/r_0,\Theta)\) plane for the NLED-Kerr black hole in AdS spacetime, with \(L=1\), \(q_m=1\), \(\beta=1\), \(a=0.05\), and \(\tau/r_0=60\). The black dot denotes the isolated zero.}
  \label{fig:jzzads_shiliangtu} 
  \end{figure}
  
Fig.~\ref{fig:jzzads_shiliangtu} illustrates the vector field for \(L=q_m=\beta=1\), \(a=0.05\), and \(\tau/r_0=60\). Equation~\eqref{eq:tau_ads} gives \(r_h/r_0=0.4367\), consistent with the black dot. The implicit-derivative argument given above proves that its Jacobian is negative. Strict monotonicity gives exactly one defect, hence \(w=W=-1\), for every admissible \(\tau\) above the finite lower endpoint.

Thus the AdS reduced model gives one unstable branch with \(W=-1\), a finite nonzero lower endpoint, and a quartic divergence at infinity. It is an explicit realization of the predicted \(\overline{W}^{1-}\) subclass throughout \(\mathcal{D}\) . 
\subsection{\label{sec:level3c}Comparison with the Kerr-AdS black hole under the same reduced prescription}
In this subsection, we briefly revisit the thermodynamic topological classification of the four-dimensional Kerr-AdS black hole as a benchmark for the NLED-Kerr result. To avoid confusion, its rotation parameter is denoted by \(a_{\mathrm{KA}}\), whereas the rotation parameter of the NLED-Kerr model remains \(a\).
Its metric in generalized Boyer-Lindquist coordinates is given by Ref.~\cite{ref49}.
\begin{equation}
\begin{aligned}
ds^2={}&-\frac{\Delta_r}{\rho^2}
\left(dt-\frac{a_{\mathrm{KA}}\sin^2\theta}{\Xi}d\phi\right)^2
+\frac{\rho^2}{\Delta_r}dr^2\\
&+\frac{\rho^2}{\Delta_\theta}d\theta^2
+\frac{\Delta_\theta\sin^2\theta}{\rho^2}\\
&\qquad\times
\left(a_{\mathrm{KA}}\,dt-\frac{r^2+a_{\mathrm{KA}}^2}{\Xi}d\phi\right)^2,
\end{aligned}
\end{equation}
where
\begin{equation}
\begin{aligned}
\Delta_r&=(r^2+a_{\mathrm{KA}}^2)\left(1+\frac{r^2}{\ell^2}\right)-2mr,\\
\Delta_\theta&=1-\frac{a_{\mathrm{KA}}^2}{\ell^2}\cos^2\theta,\\
\rho^2&=r^2+a_{\mathrm{KA}}^2\cos^2\theta,\\
\Xi&=1-\frac{a_{\mathrm{KA}}^2}{\ell^2},
\end{aligned}
\end{equation}
where \(m\) is the mass parameter, \(a_{\mathrm{KA}}\) is the rotation parameter, and \(\ell\) is the AdS radius. The parameter \(\ell\) is identical to \(L\) in the preceding subsections, and \(\Lambda=-3/\ell^2\).

The thermodynamic quantities of the Kerr-AdS black hole are well established \cite{ref49}. Restoring \(G\) explicitly in this benchmark subsection, the mass (enthalpy), angular momentum, entropy, temperature, angular velocity in the nonrotating frame at infinity, pressure, and thermodynamic volume are
\begin{align}
M_{\mathrm{KA}}&=\frac{m}{G\Xi^2},\qquad
J_{\mathrm{KA}}=\frac{a_{\mathrm{KA}}m}{G\Xi^2},\\
S_{\mathrm{KA}}&=\frac{\pi(r_h^2+a_{\mathrm{KA}}^2)}{G\Xi},\\
T_{\mathrm{KA}}&=\frac{r_h^2-a_{\mathrm{KA}}^2+\ell^{-2}(3r_h^4+a_{\mathrm{KA}}^2 r_h^2)}{4\pi r_h(r_h^2+a_{\mathrm{KA}}^2)},\\
\Omega_{\mathrm{KA}}&=\frac{a_{\mathrm{KA}}(1+r_h^2/\ell^2)}{r_h^2+a_{\mathrm{KA}}^2},\\
P&=\frac{3}{8\pi G\ell^2},\\
V_{\mathrm{KA}}&=\frac{2\pi(r_h^2+a_{\mathrm{KA}}^2)}{3r_h\Xi^2}\left(2r_h^2+a_{\mathrm{KA}}^2-\frac{a_{\mathrm{KA}}^2 r_h^2}{\ell^2}\right),
\end{align}
where $r_h$ is the event horizon radius determined by the largest root of $\Delta_r(r_h)=0$, which yields
\begin{equation}
m=\frac{(r_h^2+a_{\mathrm{KA}}^2)(1+r_h^2/\ell^2)}{2r_h}.
\end{equation}

These quantities satisfy the standard Kerr-AdS extended first law
\begin{equation}
\delta M_{\mathrm{KA}} = T_{\mathrm{KA}}\delta S_{\mathrm{KA}}+\Omega_{\mathrm{KA}}\delta J_{\mathrm{KA}}+V_{\mathrm{KA}}\delta P.
\end{equation}

For a like-for-like comparison, we impose the same fixed-rotation reduction on Kerr-AdS. The reduced off-shell functional is
\begin{equation}
\mathcal{F}_{\mathrm{KA,red}} = M_{\mathrm{KA}} - \frac{S_{\mathrm{KA}}}{\tau} - \Omega_{\mathrm{KA}}J_{\mathrm{KA}},
\end{equation}
where $\tau$ is the inverse temperature of the cavity enclosing the black hole. Substituting the Kerr-AdS thermodynamic quantities into the above expression, we obtain
\begin{equation}
\begin{aligned}
\mathcal{F}_{\mathrm{KA,red}} &= \frac{(a_{\text{KA}}^2+r_h^2)\left(1+r_h^2/\ell^2\right)}{2G\left(1-a_{\text{KA}}^2/\ell^2\right)^2 r_h}
\\
&\quad-\frac{\pi(a_{\text{KA}}^2+r_h^2)}{G\left(1-a_{\text{KA}}^2/\ell^2\right)\tau}
\\
&\quad-\frac{a_{\text{KA}}^2\left(1+r_h^2/\ell^2\right)^2}{2G\left(-1+a_{\text{KA}}^2/\ell^2\right)^2r_h}.
\end{aligned}
\end{equation}

The components of the vector $\boldsymbol{\phi}$ are then
\begin{equation}
\phi^{r_h}=\frac{\partial \mathcal{F}_{\mathrm{KA,red}}}{\partial r_h},\qquad \phi^\Theta=-\cot\Theta\csc\Theta.
\end{equation}

After simplification, the radial component takes the remarkably simple form
\begin{equation}
\phi^{r_h}=\frac{4\ell^2\pi r_h-\ell^2\tau-3r_h^2\tau}{2G(a_{\mathrm{KA}}^2-\ell^2)\,\tau}.
\end{equation}

Setting \(\phi^{r_h}=0\), we obtain the defect curve
\begin{equation}
\tau=\frac{4\ell^2\pi r_h}{\ell^2+3r_h^2}.
\end{equation}

This expression exhibits two limiting behaviors
\begin{equation}
\lim_{r_h\to 0}\tau=0,\qquad \lim_{r_h\to\infty}\tau=0.
\end{equation}

In deriving this one-dimensional curve, \(a_{\mathrm{KA}}\) is held fixed while the model expression for \(\Omega_{\mathrm{KA}}\) is inserted. It is therefore directly comparable with the NLED-Kerr reduced path, but neither curve is presented as a strict fixed-\(\Omega\) grand-canonical result.

\begin{figure}[!htbp]
  \centering
  \includegraphics[width=0.3\textwidth]{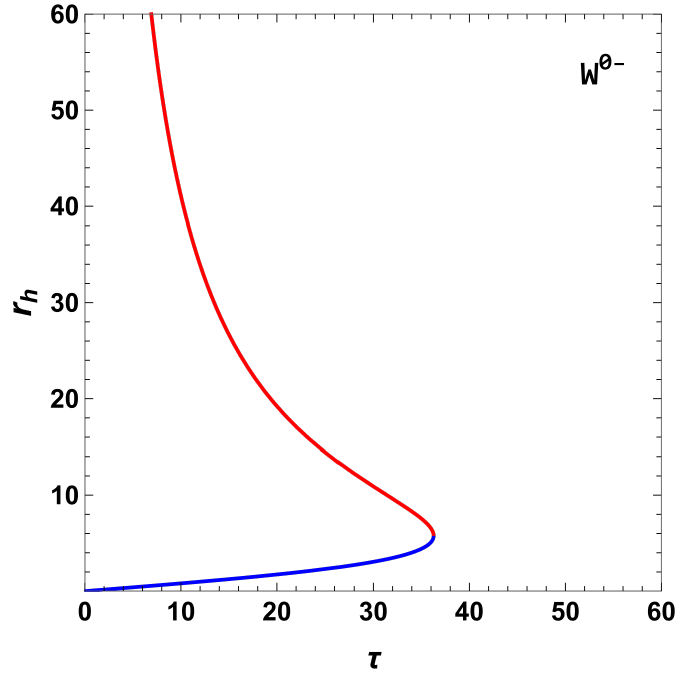}
  \caption{Inverse-temperature curve \(\tau(r_h)\) for the reduced Kerr-AdS benchmark, with \(\ell=10\) and \(a_{\mathrm{KA}}=0.05\).}
  \label{fig:jzzads_kerr} 
  \end{figure}
\begin{table*}[htbp]
\centering
\caption{Endpoint behaviors and topological statements under the reduced fixed-parameter prescription.}
\begin{tabular}{|@{}l|c|c|c@{}|c|}
\hline
Black hole model & Lower endpoint & Outer endpoint & Topological statement & $W$\\
\hline
NLED-Kerr (dS) & $\tau(0)>0$ & formal $\tau(\infty)=\infty$ & formal off-shell result & --\\\hline
NLED-Kerr (AdS) & $\tau(0)>0$ & $\tau(\infty)=\infty$ & \(\overline{W}^{1-}\) & $-1$\\\hline
Kerr-AdS & $\tau(0)=0$ & $\tau(\infty)=0$ & $W^{0-}$ & $0$\\
\hline
\end{tabular}
\label{tab:topo_compare}
\end{table*}

Fig.~\ref{fig:jzzads_kerr} displays the reduced Kerr-AdS \(r_h-\tau\) curve for \(\ell=10\) and \(a_{\mathrm{KA}}=0.05\). It has a single maximum at \(r_{\mathrm{max}}=\ell/\sqrt{3}\), \(\tau_{\mathrm{max}}=2\pi\ell/\sqrt{3}\), and vanishes as \(r_h\to0\) and \(r_h\to\infty\). For \(0<\tau<\tau_{\mathrm{max}}\), an unstable small-black-hole branch \((w=-1)\) and a stable large-black-hole branch \((w=+1)\) merge at the maximum, giving \(W=0\) and the \(W^{0-}\) classification. The AdS NLED-Kerr reduced curve instead has one unstable branch and \(W=-1\).

Thus, within the stated reduced benchmark, the two endpoint limits and alternating winding numbers place the Kerr-AdS result in the \(W^{0-}\) class. The two branches merge at an annihilation point.

This behavior contrasts with the \(\overline{W}^{1-}\) subclass identified for the NLED-Kerr model. The key distinction lies in the endpoint behavior of the inverse temperature, as summarised below.

For Kerr-AdS, the two branches have opposite winding numbers and give \(W=0\). For AdS NLED-Kerr, the finite lower endpoint, quartic divergence, and strict monotonicity give one unstable branch and \(W=-1\). 

Under the common reduced prescription, the NLED magnetic-charge model and Kerr-AdS therefore have different endpoint and branch structures. This is a comparison of two fixed-rotation paths and does not assert equivalence to a strict fixed-\(\Omega\) ensemble.

\section{Conclusions}
\label{sec:level5}

We have investigated the thermodynamic topology of the magnetically charged, slowly rotating NLED-Kerr model using an explicitly defined reduced fixed-parameter off-shell potential.

In AdS spacetime, the analytic curve is strictly increasing throughout \(\mathcal{D}\), begins at a finite nonzero value, and diverges at infinity. It contains exactly one defect with \(w=W=-1\) and therefore realizes the predicted \(\overline{W}^{1-}\) subclass \cite{ref27}. This conclusion follows analytically and is not restricted to the discrete values displayed in the figures.

In dS spacetime, the formally extended reduced curve is strictly increasing, and any isolated zero on this curve has \(w=-1\). Because the parameter-dependent physical horizon interval is not determined here, no physical dS asymptotic subclass or global topological number is assigned.

For comparison, the reduced Kerr-AdS benchmark has two branches with opposite winding numbers and \(W=0\). The NLED mass profile,
\begin{equation}
M(r)=\frac{q_m^{3/2}}{2\beta^{1/4}}\arctan\left(\frac{r}{\sqrt{q_m}\beta^{1/4}}\right),
\end{equation}
is accompanied, on the same reduced path, by a finite \(\tau(0)\), a divergent AdS outer endpoint, and the absence of the stable large-black-hole branch.

The scope is fixed by Eq.~\eqref{eq:domain}: the result concerns the resummed slow-rotation ansatz and the reduced path at fixed \(a,q_m,\beta,L\). It makes no strict grand-canonical claim and introduces no independent magnetic-potential variation. The boundaries \(a=0\), \(q_m=0\), and \(\beta=0\) are excluded reduced sectors, rather than singular limits invoked to support the classification.

\section*{Acknowledgements}
This work was supported by the National Natural Science Foundation of China (No. 12265007) and the Guizhou Provincial Major Scientific and Technological Program (XKBF(2025)010).



\bibliographystyle{elsarticle-num}
\bibliography{wenxian}

@PREAMBLE{
 "\providecommand{\noopsort}[1]{}" 
 # "\providecommand{\singleletter}[1]{#1}%" 
}

@article{ref1,
    author = "Abbott, B. P. and others",
    collaboration = "LIGO Scientific, Virgo",
    title = "{Observation of Gravitational Waves from a Binary Black Hole Merger}",
    eprint = "1602.03837",
    archivePrefix = "arXiv",
    primaryClass = "gr-qc",
    reportNumber = "LIGO-P150914",
    doi = "10.1103/PhysRevLett.116.061102",
    journal = "Phys. Rev. Lett.",
    volume = "116",
    number = "6",
    pages = "061102",
    year = "2016"
}

@article{ref2,
    author = "Witten, Edward",
    title = "{Anti de Sitter space and holography}",
    eprint = "hep-th/9802150",
    archivePrefix = "arXiv",
    reportNumber = "IASSNS-HEP-98-15",
    doi = "10.4310/ATMP.1998.v2.n2.a2",
    journal = "Adv. Theor. Math. Phys.",
    volume = "2",
    pages = "253--291",
    year = "1998"
}

@article{ref3,
    author = "Hawking, S. W.",
    editor = "Gibbons, G. W. and Hawking, S. W.",
    title = "{Particle Creation by Black Holes}",
    doi = "10.1007/BF02345020",
    journal = "Commun. Math. Phys.",
    volume = "43",
    pages = "199--220",
    year = "1975",
    note = "[Erratum: Commun.Math.Phys. 46, 206 (1976)]"
}

@article{ref4,
    author = "Bekenstein, Jacob D.",
    title = "{Black holes and entropy}",
    doi = "10.1103/PhysRevD.7.2333",
    journal = "Phys. Rev. D",
    volume = "7",
    pages = "2333--2346",
    year = "1973"
}

@article{ref7,
    author = "Frassino, Antonia M. and Kubiznak, David and Mann, Robert B. and Simovic, Fil",
    title = "{Multiple Reentrant Phase Transitions and Triple Points in Lovelock Thermodynamics}",
    eprint = "1406.7015",
    archivePrefix = "arXiv",
    primaryClass = "hep-th",
    doi = "10.1007/JHEP09(2014)080",
    journal = "JHEP",
    volume = "09",
    pages = "080",
    year = "2014"
}

@article{ref8,
    author = "Altamirano, Natacha and Kubiz{\v{n}}{\'a}k, David and Mann, Robert B. and Sherkatghanad, Zeinab",
    title = "{Kerr-AdS analogue of triple point and solid/liquid/gas phase transition}",
    eprint = "1308.2672",
    archivePrefix = "arXiv",
    primaryClass = "hep-th",
    doi = "10.1088/0264-9381/31/4/042001",
    journal = "Class. Quant. Grav.",
    volume = "31",
    pages = "042001",
    year = "2014"
}

@article{ref9,
    author = "Wei, Shao-Wen and Liu, Yu-Xiao",
    title = "{Critical phenomena and thermodynamic geometry of charged Gauss-Bonnet AdS black holes}",
    eprint = "1209.1707",
    archivePrefix = "arXiv",
    primaryClass = "gr-qc",
    doi = "10.1103/PhysRevD.87.044014",
    journal = "Phys. Rev. D",
    volume = "87",
    number = "4",
    pages = "044014",
    year = "2013"
}

@article{ref10,
    author = "Wei, Shao-Wen and Liu, Yu-Xiao and Mann, Robert B.",
    title = "{Repulsive Interactions and Universal Properties of Charged Anti{\textendash}de Sitter Black Hole Microstructures}",
    eprint = "1906.10840",
    archivePrefix = "arXiv",
    primaryClass = "gr-qc",
    doi = "10.1103/PhysRevLett.123.071103",
    journal = "Phys. Rev. Lett.",
    volume = "123",
    number = "7",
    pages = "071103",
    year = "2019"
}

@article{ref11,
    author = "Caceres, Elena and Nguyen, Phuc H. and Pedraza, Juan F.",
    title = "{Holographic entanglement entropy and the extended phase structure of STU black holes}",
    eprint = "1507.06069",
    archivePrefix = "arXiv",
    primaryClass = "hep-th",
    reportNumber = "UTTG-14-15, TCC-005-15",
    doi = "10.1007/JHEP09(2015)184",
    journal = "JHEP",
    volume = "09",
    pages = "184",
    year = "2015"
}

@article{ref12,
    author = "Hawking, S. W. and Page, Don N.",
    title = "{Thermodynamics of Black Holes in anti-De Sitter Space}",
    reportNumber = "PRINT-83-0019 (CAMBRIDGE)",
    doi = "10.1007/BF01208266",
    journal = "Commun. Math. Phys.",
    volume = "87",
    pages = "577",
    year = "1983"
}

@article{ref13,
    author = "Kubiznak, David and Mann, Robert B.",
    title = "{P-V criticality of charged AdS black holes}",
    eprint = "1205.0559",
    archivePrefix = "arXiv",
    primaryClass = "hep-th",
    doi = "10.1007/JHEP07(2012)033",
    journal = "JHEP",
    volume = "07",
    pages = "033",
    year = "2012"
}

@article{ref14,
    author = "Wei, Shao-Wen and Liu, Yu-Xiao",
    title = "{Topology of black hole thermodynamics}",
    eprint = "2112.01706",
    archivePrefix = "arXiv",
    primaryClass = "gr-qc",
    doi = "10.1103/PhysRevD.105.104003",
    journal = "Phys. Rev. D",
    volume = "105",
    number = "10",
    pages = "104003",
    year = "2022"
}

@article{ref15,
    author = "Wei, Shao-Wen and Liu, Yu-Xiao and Mann, Robert B.",
    title = "{Universal topological classifications of black hole thermodynamics}",
    eprint = "2409.09333",
    archivePrefix = "arXiv",
    primaryClass = "gr-qc",
    doi = "10.1103/PhysRevD.110.L081501",
    journal = "Phys. Rev. D",
    volume = "110",
    number = "8",
    pages = "L081501",
    year = "2024"
}

@article{ref16,
    author = "Bai, Ning-Chen and Li, Lei and Tao, Jun",
    title = "{Topology of black hole thermodynamics in Lovelock gravity}",
    eprint = "2208.10177",
    archivePrefix = "arXiv",
    primaryClass = "gr-qc",
    doi = "10.1103/PhysRevD.107.064015",
    journal = "Phys. Rev. D",
    volume = "107",
    number = "6",
    pages = "064015",
    year = "2023"
}

@article{ref17,
    author = "Alipour, Mohammad Reza and Afshar, Mohammad Ali S. and Noori Gashti, Saeed and Sadeghi, Jafar",
    title = "{Topological classification and black hole thermodynamics}",
    eprint = "2305.05595",
    archivePrefix = "arXiv",
    primaryClass = "gr-qc",
    doi = "10.1016/j.dark.2023.101361",
    journal = "Phys. Dark Univ.",
    volume = "42",
    pages = "101361",
    year = "2023"
}

@article{ref18,
    author = "Sadeghi, Jafar and Noori Gashti, Saeed and Alipour, Mohammad Reza and Afshar, Mohammad Ali S.",
    title = "{Bardeen black hole thermodynamics from topological perspective}",
    eprint = "2306.05692",
    archivePrefix = "arXiv",
    primaryClass = "hep-th",
    doi = "10.1016/j.aop.2023.169391",
    journal = "Annals Phys.",
    volume = "455",
    pages = "169391",
    year = "2023"
}

@article{ref19,
    author = "Wan, Yu-Die and Zhao, Peng and Long, Zheng-Wen",
    title = "{Novel topological subclass in Bardeen-AdS-class black holes}",
    eprint = "2607.14173",
    archivePrefix = "arXiv",
    primaryClass = "gr-qc",
    month = "7",
    year = "2026"
}

@article{ref20,
    author = "Wan, Yu-Die and Zhao, Peng and Zhang, Meng-Yao and Long, Zheng-Wen",
    title = "{Universal Thermodynamic Topological Classes of BTZ Black Holes in Einstein and F(R) Gravity}",
    eprint = "2605.26481",
    archivePrefix = "arXiv",
    primaryClass = "gr-qc",
    month = "5",
    year = "2026"
}

@article{ref21,
    author = "Wei, Shao-Wen and Liu, Yu-Xiao",
    title = "{Topology of black hole thermodynamics: A brief review}",
    eprint = "2605.00037",
    archivePrefix = "arXiv",
    primaryClass = "gr-qc",
    doi = "10.1007/s11433-025-2923-3",
    journal = "Sci. China Phys. Mech. Astron.",
    volume = "69",
    number = "6",
    pages = "260401",
    year = "2026"
}

@article{ref22,
    author = "Yerra, Pavan Kumar and Bhamidipati, Chandrasekhar",
    title = "{Topology of black hole thermodynamics in Gauss-Bonnet gravity}",
    eprint = "2202.10288",
    archivePrefix = "arXiv",
    primaryClass = "gr-qc",
    doi = "10.1103/PhysRevD.105.104053",
    journal = "Phys. Rev. D",
    volume = "105",
    number = "10",
    pages = "104053",
    year = "2022"
}

@article{ref23,
    author = "Wu, Di",
    title = "{Consistent thermodynamics and topological classes for the four-dimensional Lorentzian charged Taub-NUT spacetimes}",
    eprint = "2306.02324",
    archivePrefix = "arXiv",
    primaryClass = "gr-qc",
    doi = "10.1140/epjc/s10052-023-11782-7",
    journal = "Eur. Phys. J. C",
    volume = "83",
    number = "7",
    pages = "589",
    year = "2023"
}

@article{ref24,
    author = "Fan, Zhong-Ying",
    title = "{Topological interpretation for phase transitions of black holes}",
    eprint = "2211.12957",
    archivePrefix = "arXiv",
    primaryClass = "gr-qc",
    doi = "10.1103/PhysRevD.107.044026",
    journal = "Phys. Rev. D",
    volume = "107",
    number = "4",
    pages = "044026",
    year = "2023"
}

@article{ref25,
    author = "Sadeghi, J. and Afshar, M. A S. and Noori Gashti, S. and Alipour, M. R.",
    title = "{Topology of Hayward-AdS black hole thermodynamics}",
    doi = "10.1088/1402-4896/ad186b",
    journal = "Phys. Scripta",
    volume = "99",
    number = "2",
    pages = "025003",
    year = "2024"
}

@article{ref26,
    author = "Wu, Shan-Ping and Wei, Shao-Wen",
    title = "{Thermodynamical topology of quantum BTZ black hole}",
    eprint = "2403.14167",
    archivePrefix = "arXiv",
    primaryClass = "gr-qc",
    doi = "10.1103/PhysRevD.110.024054",
    journal = "Phys. Rev. D",
    volume = "110",
    number = "2",
    pages = "024054",
    year = "2024"
}

@article{ref27,
    author = "Wu, Di and Liu, Wentao and Wu, Shuang-Qing and Mann, Robert B.",
    title = "{Novel topological classes in black hole thermodynamics}",
    eprint = "2411.10102",
    archivePrefix = "arXiv",
    primaryClass = "hep-th",
    doi = "10.1103/PhysRevD.111.L061501",
    journal = "Phys. Rev. D",
    volume = "111",
    number = "6",
    pages = "L061501",
    year = "2025"
}

@article{ref28,
    author = "Ai, Wangyu and Wu, Di",
    title = "{W{\textasciitilde}1+ subclass: Extending the topological classification of black hole thermodynamics}",
    eprint = "2509.03308",
    archivePrefix = "arXiv",
    primaryClass = "hep-th",
    doi = "10.1103/dy7y-j24r",
    journal = "Phys. Rev. D",
    volume = "112",
    number = "12",
    pages = "124024",
    year = "2025"
}

@article{ref29,
    author = "Ayon-Beato, Eloy and Garcia, Alberto",
    title = "{Regular black hole in general relativity coupled to nonlinear electrodynamics}",
    eprint = "gr-qc/9911046",
    archivePrefix = "arXiv",
    doi = "10.1103/PhysRevLett.80.5056",
    journal = "Phys. Rev. Lett.",
    volume = "80",
    pages = "5056--5059",
    year = "1998"
}

@article{ref30,
    author = "Ayon-Beato, Eloy and Garcia, Alberto",
    title = "{The Bardeen model as a nonlinear magnetic monopole}",
    eprint = "gr-qc/0009077",
    archivePrefix = "arXiv",
    doi = "10.1016/S0370-2693(00)01125-4",
    journal = "Phys. Lett. B",
    volume = "493",
    pages = "149--152",
    year = "2000"
}

@article{ref31,
    author = "Bronnikov, Kirill A.",
    title = "{Regular magnetic black holes and monopoles from nonlinear electrodynamics}",
    eprint = "gr-qc/0006014",
    archivePrefix = "arXiv",
    doi = "10.1103/PhysRevD.63.044005",
    journal = "Phys. Rev. D",
    volume = "63",
    pages = "044005",
    year = "2001"
}

@article{ref32,
    author = "Ayon-Beato, Eloy and Garcia, Alberto",
    title = "{New regular black hole solution from nonlinear electrodynamics}",
    eprint = "hep-th/9911174",
    archivePrefix = "arXiv",
    doi = "10.1016/S0370-2693(99)01038-2",
    journal = "Phys. Lett. B",
    volume = "464",
    pages = "25",
    year = "1999"
}

@article{ref33,
    author = "Fan, Zhong-Ying and Wang, Xiaobao",
    title = "{Construction of Regular Black Holes in General Relativity}",
    eprint = "1610.02636",
    archivePrefix = "arXiv",
    primaryClass = "gr-qc",
    doi = "10.1103/PhysRevD.94.124027",
    journal = "Phys. Rev. D",
    volume = "94",
    number = "12",
    pages = "124027",
    year = "2016"
}

@article{ref34,
    author = "Born, M. and Infeld, L.",
    title = "{Foundations of the new field theory}",
    doi = "10.1098/rspa.1934.0059",
    journal = "Proc. Roy. Soc. Lond. A",
    volume = "144",
    number = "852",
    pages = "425--451",
    year = "1934"
}

@article{ref36,
    author = "Balart, Leonardo and Vagenas, Elias C.",
    title = "{Regular black holes with a nonlinear electrodynamics source}",
    eprint = "1408.0306",
    archivePrefix = "arXiv",
    primaryClass = "gr-qc",
    doi = "10.1103/PhysRevD.90.124045",
    journal = "Phys. Rev. D",
    volume = "90",
    number = "12",
    pages = "124045",
    year = "2014"
}

@article{ref38,
    author = "Yokoyama, Shuichiro and Suyama, Teruaki and Tanaka, Takahiro",
    title = "{Primordial Non-Gaussianity in Multi-Scalar Inflation}",
    eprint = "0711.2920",
    archivePrefix = "arXiv",
    primaryClass = "astro-ph",
    reportNumber = "KUNS-2110",
    doi = "10.1103/PhysRevD.77.083511",
    journal = "Phys. Rev. D",
    volume = "77",
    pages = "083511",
    year = "2008"
}

@article{ref39,
    author = "Choi, Ki-Young and Seto, Osamu",
    title = "{A Dirac right-handed sneutrino dark matter and its signature in the gamma-ray lines}",
    eprint = "1205.3276",
    archivePrefix = "arXiv",
    primaryClass = "hep-ph",
    reportNumber = "APCTP-PRE2012-008, HGU-CAP-017",
    doi = "10.1103/PhysRevD.86.043515",
    journal = "Phys. Rev. D",
    volume = "86",
    pages = "043515",
    year = "2012",
    note = "[Erratum: Phys.Rev.D 86, 089904 (2012)]"
}

@article{ref40,
    author = "Kruglov, S. I.",
    title = "{Black hole as a magnetic monopole within exponential nonlinear electrodynamics}",
    eprint = "1703.02029",
    archivePrefix = "arXiv",
    primaryClass = "gr-qc",
    doi = "10.1016/j.aop.2016.12.036",
    journal = "Annals Phys.",
    volume = "378",
    pages = "59--70",
    year = "2017"
}

@article{ref41,
    author = "Wei, Shao-Wen and Liu, Yu-Xiao and Mann, Robert B.",
    title = "{Black Hole Solutions as Topological Thermodynamic Defects}",
    eprint = "2208.01932",
    archivePrefix = "arXiv",
    primaryClass = "gr-qc",
    doi = "10.1103/PhysRevLett.129.191101",
    journal = "Phys. Rev. Lett.",
    volume = "129",
    number = "19",
    pages = "191101",
    year = "2022"
}

@article{ref42,
    author = "Gibbons, G. W. and Hawking, S. W.",
    title = "{Action Integrals and Partition Functions in Quantum Gravity}",
    reportNumber = "PRINT-76-0995 (CAMBRIDGE)",
    doi = "10.1103/PhysRevD.15.2752",
    journal = "Phys. Rev. D",
    volume = "15",
    pages = "2752--2756",
    year = "1977"
}

@article{ref43,
  title = "{Black-hole thermodynamics and the Euclidean Einstein action}",
  author = "{York, James W.}",
  journal = "{Phys. Rev. D}",
  volume = "{33}",
  issue = "{8}",
  pages = "{2092--2099}",
  numpages = "{0}",
  year = "{1986}",
  month = "{Apr}",
  publisher = "{American Physical Society}",
  doi = "{10.1103/PhysRevD.33.2092}",
}

@article{ref44,
  title = {Generalized free energy landscape of a black hole phase transition},
  author = {Li, Ran and Wang, Jin},
  journal = {Phys. Rev. D},
  volume = {106},
  issue = {10},
  pages = {106015},
  numpages = {12},
  year = {2022},
  month = {Nov},
  publisher = {American Physical Society},
  doi = {10.1103/PhysRevD.106.106015},
 
}

@article{ref45,
  title = {Universal topological classifications of black hole thermodynamics},
  author = {Wei, Shao-Wen and Liu, Yu-Xiao and Mann, Robert B.},
  journal = {Phys. Rev. D},
  volume = {110},
  issue = {8},
  pages = {L081501},
  numpages = {6},
  year = {2024},
  month = {Oct},
  publisher = {American Physical Society},
  doi = {10.1103/PhysRevD.110.L081501},
  
}

@article{ref46,
    author = "Duan, Yishi",
    title = "{THE STRUCTURE OF THE TOPOLOGICAL CURRENT}",
    reportNumber = "SLAC-PUB-3301",
    month = "3",
    year = "1984"
}

@article{ref47,
    author = "Chen, Hao and Zhang, Meng-Yao and Hassanabadi, Hassan and Huang, Qihong and Long, Zheng-Wen",
    title = "{Novel topological subclass in Ho{\v r}ava{\textendash}Lifshitz black holes}",
    eprint = "2511.01367",
    archivePrefix = "arXiv",
    primaryClass = "gr-qc",
    doi = "10.1140/epjc/s10052-025-15133-6",
    journal = "Eur. Phys. J. C",
    volume = "85",
    number = "12",
    pages = "1386",
    year = "2025"
}

@article{ref48,
    author = "Pawar, Vinayak S. and Das, Siba Prasad",
    title = "{Thermodynamic properties of the Kerr Black-hole in non-linear electrodynamics with cosmological constant}",
    eprint = "2604.26586",
    archivePrefix = "arXiv",
    primaryClass = "gr-qc",
    month = "4",
    year = "2026"
}

@article{ref49,
    author = "Xu, Yingnan and Chu, Shuangshuang",
    title = "{Extended Hamiltonian thermodynamics and Carroll contractions of AdS black holes}",
    eprint = "2604.27449",
    archivePrefix = "arXiv",
    primaryClass = "hep-th",
    month = "4",
    year = "2026"
}

@article{refa,
    author = "Eslam Panah, Behzad and Hamil, Bilel and Rodrigues, Manuel E.",
    title = "{Born{\textendash}Infeld AdS black holes surrounded by perfect fluid dark matter}",
    eprint = "2603.14159",
    archivePrefix = "arXiv",
    primaryClass = "gr-qc",
    doi = "10.1140/epjc/s10052-026-15823-9",
    journal = "Eur. Phys. J. C",
    volume = "86",
    number = "6",
    pages = "632",
    year = "2026"
}

@article{refb1,
    author = "Hazarika, Bidyut and Eslam Panah, B. and Phukon, Prabwal",
    title = "{Thermodynamic topology of topological charged dilatonic black holes}",
    eprint = "2407.05325",
    archivePrefix = "arXiv",
    primaryClass = "hep-th",
    doi = "10.1140/epjc/s10052-024-13598-5",
    journal = "Eur. Phys. J. C",
    volume = "84",
    number = "11",
    pages = "1204",
    year = "2024"
}

@article{refb2,
    author = "Eslam Panah, B. and Hazarika, B. and Phukon, P.",
    title = "{Thermodynamic Topology of Topological Black Hole in F(R)-ModMax Gravity{\textquoteright}s Rainbow}",
    eprint = "2405.20022",
    archivePrefix = "arXiv",
    primaryClass = "hep-th",
    doi = "10.1093/ptep/ptae116",
    journal = "PTEP",
    volume = "2024",
    number = "8",
    pages = "083E02",
    year = "2024"
}

@article{refb3,
    author = "Jeon, Imtak and Lee, Bum-Hoon and Lee, Wonwoo and Mishra, Madhu",
    title = "{Stability and topological nature of charged Gauss-Bonnet AdS black holes in five dimensions}",
    eprint = "2407.20016",
    archivePrefix = "arXiv",
    primaryClass = "hep-th",
    doi = "10.1103/PhysRevD.111.064006",
    journal = "Phys. Rev. D",
    volume = "111",
    number = "6",
    pages = "064006",
    year = "2025"
}

@article{refc,
    author = "Jiang, Hong-Yue and Liu, Yu-Xiao",
    title = "{Complexity equals anything for multi-horizon black holes}",
    eprint = "2506.10398",
    archivePrefix = "arXiv",
    primaryClass = "hep-th",
    doi = "10.1007/JHEP12(2025)072",
    journal = "JHEP",
    volume = "12",
    pages = "072",
    year = "2025"
}

@article{refd1,
    author = "Fang, Chaoxi and Jiang, Jie and Zhang, Ming",
    title = "{Revisiting thermodynamic topologies of black holes}",
    eprint = "2211.15534",
    archivePrefix = "arXiv",
    primaryClass = "gr-qc",
    doi = "10.1007/JHEP01(2023)102",
    journal = "JHEP",
    volume = "01",
    pages = "102",
    year = "2023"
}

@article{refd2,
    author = "Hazarika, Bidyut and Phukon, Prabwal",
    title = "{Thermodynamic Topology of Black Holes in f(R) Gravity}",
    eprint = "2401.16756",
    archivePrefix = "arXiv",
    primaryClass = "hep-th",
    doi = "10.1093/ptep/ptae035",
    journal = "PTEP",
    volume = "2024",
    number = "4",
    pages = "043E01",
    year = "2024"
}

@article{refd3,
    author = "Zhu, Xiao-Dan and Wu, Di and Wen, Dan",
    title = "{Topological classes of thermodynamics of the rotating charged AdS black holes in gauged supergravities}",
    eprint = "2402.15531",
    archivePrefix = "arXiv",
    primaryClass = "hep-th",
    doi = "10.1016/j.physletb.2024.138919",
    journal = "Phys. Lett. B",
    volume = "856",
    pages = "138919",
    year = "2024"
}

@article{refd4,
    author = "Zhu, Xiao-Dan and Liu, Wentao and Wu, Di",
    title = "{Universal thermodynamic topological classes of rotating black holes}",
    eprint = "2409.12747",
    archivePrefix = "arXiv",
    primaryClass = "hep-th",
    doi = "10.1016/j.physletb.2024.139163",
    journal = "Phys. Lett. B",
    volume = "860",
    pages = "139163",
    year = "2025"
}

@article{refDRGTString2026,
  author  = {Chen, Hao and Zhang, Meng-Yao and Yang, Yi and Huang, Qihong and Long, Zheng-Wen},
  title   = {Universal thermodynamic topological classes of the charged {dRGT} black string},
  journal = {Phys. Lett. B},
  volume  = {880},
  pages   = {140731},
  year    = {2026},
  doi     = {10.1016/j.physletb.2026.140731},
  eprint  = {2605.28523},
  archivePrefix = {arXiv}
}

@article{refCKG2025,
  author  = {Chen, Hao and Wu, Di and Zhang, Meng-Yao and Zare, Soroush and Hassanabadi, Hassan and L{\"u}tf{\"u}o\u{g}lu, Bekir Can and Long, Zheng-Wen},
  title   = {Universal thermodynamic topological classes of static black holes in {Conformal Killing Gravity}},
  journal = {Eur. Phys. J. C},
  volume  = {85},
  number  = {8},
  pages   = {828},
  year    = {2025},
  doi     = {10.1140/epjc/s10052-025-14581-4},
  eprint  = {2506.03695},
  archivePrefix = {arXiv}
}

@article{refTensorAdS2026,
  author  = {Chen, Hao and Zhang, Meng-Yao and Ara{\'u}jo Filho, A. A. and Hosseinifar, F. and Hassanabadi, Hassan},
  title   = {Thermal, topological, and scattering effects of an {AdS} charged black hole with an antisymmetric tensor background},
  journal = {Commun. Theor. Phys.},
  volume  = {78},
  number  = {2},
  pages   = {025404},
  year    = {2026},
  doi     = {10.1088/1572-9494/ae0393},
  eprint  = {2408.03090},
  archivePrefix = {arXiv}
}

@article{refPhantomMassive2024,
  author  = {Chen, Hao and Wu, Di and Zhang, Meng-Yao and Hassanabadi, Hassan and Long, Zheng-Wen},
  title   = {Thermodynamic topology of phantom {AdS} black holes in massive gravity},
  journal = {Phys. Dark Univ.},
  volume  = {46},
  pages   = {101617},
  year    = {2024},
  doi     = {10.1016/j.dark.2024.101617},
  eprint  = {2404.08243},
  archivePrefix = {arXiv}
}

\end{document}